\documentclass[conference]{IEEEtran}
\IEEEoverridecommandlockouts

\usepackage[utf8]{inputenc}
\usepackage[T1]{fontenc}
\usepackage{cite}
\usepackage{amsmath,amssymb,amsfonts}
\usepackage{graphicx}
\usepackage{textcomp}
\usepackage{xcolor}
\usepackage{colortbl}
\usepackage{booktabs}
\usepackage{tabularx}
\usepackage{xurl}
\usepackage[hidelinks]{hyperref}
\usepackage{lipsum}

\begin{document}

\title{Is This Code Mine? Student Ownership\\When AI Agents Write the Spec and the Code}

\author{\IEEEauthorblockN{Anonymous Author(s)}
\IEEEauthorblockA{\textit{Anonymous Institution(s)}}
}
% Camera-ready:
\author{\IEEEauthorblockN{Gustavo Pinto}
\IEEEauthorblockA{\textit{Federal University of Pará (UFPA)}\\
Belém, Brazil \\
gpinto@ufpa.br}
\and
\IEEEauthorblockN{Enne Rebeca Silva de Freitas}
\IEEEauthorblockA{\textit{Federal University of Pará (UFPA)}\\
Belém, Brazil \\
enne.freitas@icen.ufpa.br}
\and
\IEEEauthorblockN{Danilo Monteiro Ribeiro}
\IEEEauthorblockA{\textit{CESAR School}\\
Recife, Brazil \\
dmr@cesar.school}
}

\maketitle

\begin{abstract}
Software engineering courses are starting to ask students to build entire systems with AI coding agents, but little is known about what students actually do with them and whether they still own what they deliver. If the agent writes both the code and the decisions behind it, the artifacts we grade no longer show what students learned. We studied an assignment in which 18 pairs of students built applications with agents under Spec-Driven Development, combining the mined history of their repositories, the prompts they recorded, and a survey of 33 students. The agents wrote 69\% of the specifications meant to keep students in control, sometimes from a prompt of a few words. Students locate their ownership in the ideas, crediting themselves with half of them but a fifth of the code, yet most could not explain that code to a colleague or modify it without the agent, and what they report having written is unrelated to what their own commits declare. The three teams that met the 100,000-line target did so with generated code the application never uses, in one case proposed by the agent itself, and one team rewrote its history to backdate its specifications. Teaching with agents should therefore assess what students specify, review, and can explain, rather than what they ship, and enforce the engineering process with tools rather than trusting it to students or agents.
\end{abstract}

\begin{IEEEkeywords}
AI coding agents, spec-driven development, software engineering education, psychological ownership, mining software repositories
\end{IEEEkeywords}

\section{Introduction}

Today's students will graduate into a profession in which AI coding agents read a code base, plan a change, edit dozens of files, run the tests, and commit the result~\cite{li2026aidev}. Agentic tools are already listed in the syllabi of software engineering courses~\cite{geng2026mapping}, and instructors are rethinking assignments, assessment, and learning goals for a world in which generating code is cheap~\cite{prather2023,denny2024,lau2023ban}. A pair of students can now generate in a few days a volume of code that once required a team and a semester. What they learn by doing so, and what they lose, is an open question for software engineering education.

Research on students and AI, however, has largely focused on introductory programming, where tasks tend to be small and a central concern is whether the tool helps or hinders novices learning to code~\cite{kazemitabaar2023,prather2024widening,prather2023}. Studies in software engineering courses are more recent and have mostly examined chat assistants supporting parts of a team project~\cite{baresi2025students,salomon2025exploration,rasnayaka2024empirical}, or novices vibe coding small programs~\cite{geng2026vibe,gama2026vibes}. Much less is known about what happens at the other end of the curriculum, when students are asked to use agents on purpose, at a scale they could never reach by hand. At this scale, the question is no longer whether the student wrote the code, but whether the student is still its owner: whether they made the decisions it embodies, know where things are, and feel responsible for its defects. Studies on AI-assisted writing suggest that this feeling may not survive: people do not perceive ownership of AI-generated text, yet still present themselves as its authors~\cite{draxler2024}. If the same happens with code, we may be training engineers who ship systems they do not feel are theirs, and do not know how to maintain.

Spec-Driven Development (SDD) is one of the answers practitioners have proposed to keep humans in control of agents~\cite{piskala2026sdd}, supported by tools such as GitHub Spec Kit~\cite{speckit} and Kiro~\cite{kiro}. Before asking for code, the developer writes a specification with verifiable acceptance criteria, then a plan, then a list of tasks, and the agent implements one task at a time against the specification. In SDD, the specification is the artifact the human owns, and the code is its consequence. Whether students actually work this way under deadline pressure, and whether owning the specification translates into owning the code, are open questions.

In this paper, we study these questions in a real course. In the first assignment of an elective course on AI-assisted software development at a university in Latin America, 18 pairs of students had to deliver a published application with at least 100,000 lines of code, built with AI coding agents under SDD. Every specification had to be dated and committed before the code it describes, every commit had to declare the agent that produced it and the specification it implements, and every prompt had to be recorded in the repository. These rules leave a detailed trace of how the code was produced, which we mine and combine with a survey of the students' perceived ownership. We answer five research questions:

\begin{itemize}
\item[\textbf{RQ1}] \emph{To what extent do students follow SDD when working with agents?} We check each repository against the rules of the assignment, such as publishing the application, writing the specification first, and recording every session.
\item[\textbf{RQ2}] \emph{How do students specify software for AI coding agents?} SDD assigns the specification to the human and the code to the agent. We analyze the 277 specifications delivered to find out who wrote them, whether they have the structure the method requires, and whether they actually govern the code.
\item[\textbf{RQ3}] \emph{How much ownership do students perceive over the code, the specification, and the application they delivered?} We survey the students on their perceived ownership, their knowledge of the code, and the share of the ideas and of the code they attribute to themselves and to the agent.
\item[\textbf{RQ4}] \emph{How does perceived ownership relate to the agent's involvement as recorded in the repository?} We relate the answers of each student to the data mined from their repository, compare what they report having written with what their own commits declare, and read their open answers against their repositories.
\item[\textbf{RQ5}] \emph{Do students game the requirements of the assignment, and what role do AI agents play?} We search the repositories for signs that requirements were met in letter but not in spirit, such as code that counts toward the target but is never used, or a process record that does not match what happened, and trace each case to the commits and sessions behind it to understand the part the agents played.
\end{itemize}

Students followed the rules unevenly: three teams recorded no session, and
the recorded sessions cover only about half of the commits made by agents
(RQ1). Agents committed 69\% of the specifications, sometimes from a prompt
of a few words; one in seven entered the repository with the code it
describes, and only 39\% of the added lines can be traced to a specification
(RQ2). Students credit themselves with half of the ideas but a fifth of the
code, and most could neither explain nor modify it without the agent (RQ3).
Ownership does not track the agent's recorded involvement, and what students
report having written shows no association with what their commits declare;
it tracks instead how much of the generated code they read (RQ4). Finally,
the three projects that reached the size target did so with generated code
the application never uses, in one case proposed by the agent, and one team
rewrote its history to backdate its specifications (RQ5).
% This paper makes the following contributions:
% \begin{itemize}
% \item A description of an assignment that asks students to build, with AI coding agents, a system larger than any they could write by hand, with rules that make the use of the agent traceable;
% \item An empirical characterization of 18 applications, with 780,614 lines of code and 1,051 commits, and of how closely students followed SDD;
% \item An analysis of how students gamed the requirements of the assignment, of the role AI agents played, and of what made the gaming visible;
% \item An analysis of 277 specifications written for AI coding agents, showing that the agent wrote most of the artifact that SDD assigns to the human, and that specifications only partially govern the code;
% \item The first study, to the best of our knowledge, of students' perceived ownership of code written with AI coding agents, combining a survey with the mined history of their repositories;
% \item \textcolor{red}{An artifact folder with the questionnaire, the anonymized responses, the list of repositories, and the scripts and code used in the analyses (Section~\ref{sec:data}).}
% \end{itemize}

\section{Methodology}

\subsection{Course and Assignment}

\emph{AI-Assisted Software Development} is an elective course on building and delivering software systems with AI coding agents as the main working tool: students are expected to specify, direct, review, and test what the agents produce, and to answer for it. The course runs for one semester, with two meetings per week (60 hours in total), and had 48 students enrolled in 2026. Its first part has ten lectures, covering, among other topics, how coding agents work, context engineering, SDD, the review of generated code, skills, the Model Context Protocol (MCP), and loop engineering; the second part is a project assessed in milestones and in an individual oral defense. All activities are done in pairs, and the use of agents is mandatory.

The first assignment, released after the lecture on SDD, asked each pair to choose and build a non-trivial application with an AI coding agent, following the \emph{spec, plan, tasks} workflow presented in class; students were free to choose the domain, the stack, and the tools. It was released on September 29, 2026, students had one week, until October 6, and it accounted for 30\% of the final grade. It had four requirements: the application had to open at a public URL; it had to have at least 100,000 lines of code, counted with \texttt{cloc} excluding dependencies, lock files, build artifacts, documentation, and data; each part of the system needed a dated specification in a \texttt{SPEC/} folder, written before the code it describes; and every prompt had to be recorded, with the tool and the model.

Conventions made the process traceable. Specifications were named \texttt{YYYY-MM-DD-<part>.md} and had to state what is built and why, list verifiable acceptance criteria, and state what is out of scope; a change of direction required a new dated file stating which one it replaces. The raw export of every session went into \texttt{prompts/sessoes/} (\emph{sessions}). Every commit declared its origin in two trailers, \texttt{Agent:} (e.g., \texttt{claude-code/claude-sonnet-5-5}, \texttt{none} for handwritten code, or \texttt{+~manual} for generated and hand-edited code) and \texttt{Spec:}, pointing to the specification it implements. Both members had to commit with their own accounts, since the commit author identifies who drove each session, and history could not be rewritten.

\subsection{Data Collection}

On October 7, 2026, we cloned the full history of the 18 repositories and read their \texttt{README.md} files to identify the production URL and the tools. We extracted the prompts from the recorded sessions, with a parser for the export format of each tool; 87 of the 109 session files could be parsed, and the others are copies of the chat interface that do not mark who is speaking.

\begin{figure*}[!t]
\centering
\includegraphics[width=\textwidth]{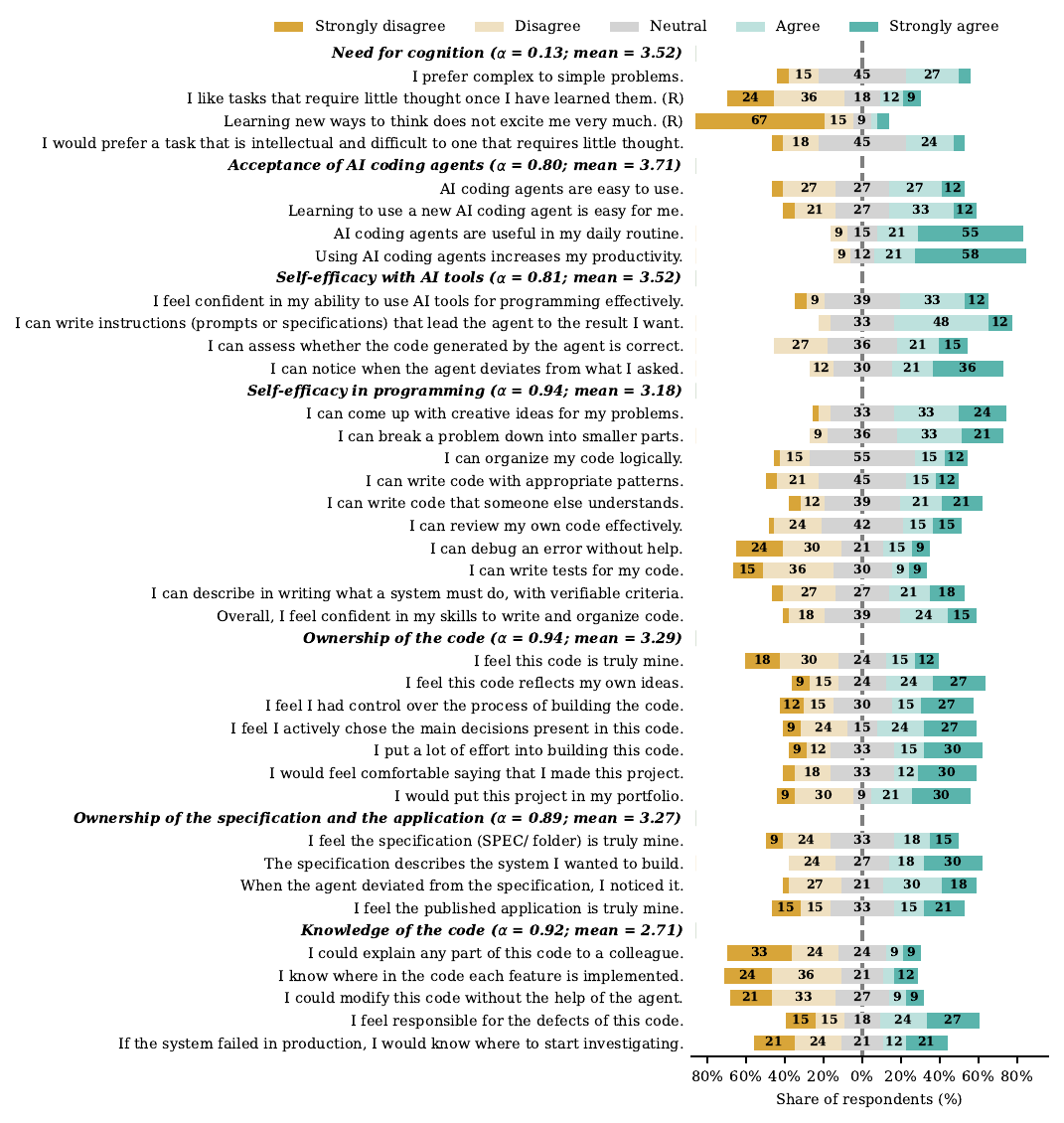}
\caption{Answers of the 33 respondents to the Likert statements of the ownership survey, grouped by construct: share of respondents at each point of the scale (labels below 8\% omitted). (R) marks reverse-coded statements; construct means use inverted values. The statements on need for cognition do not form a construct ($\alpha = 0.13$) and are not used in the analyses.}
\label{fig:survey}
\end{figure*}

On the same day, after the presentations, we invited the 36 students who delivered the assignment to an individual survey, following the phases of Kitchenham and Pfleeger~\cite{kitchenham2008}; we analyze the answers received until October 13. The questionnaire, in Google Forms, was adapted from the one used by Gero et al.\ to study ownership in AI-assisted writing~\cite{gero2026planning}, and three researchers answered it and suggested improvements before it was sent. It has 28 questions in six sections, on demographics, prior experience with programming and AI, the share of the ideas and of the code that respondents attribute to themselves and to the agent, the share of the generated code they read, and the hours spent, plus five optional open questions, such as why they felt, or did not feel, that the code was theirs. Seven groups of statements, 38 in total, are answered on a five-point Likert scale (Figure~\ref{fig:survey}): need for cognition~\cite{cacioppo1982}, acceptance of AI coding agents~\cite{davis1989}, self-efficacy with AI tools and in programming, and the perceived ownership~\cite{pierce2001} of the code and of the specification and application, and knowledge of the code. The form asked for the GitHub username of the respondent, which we use to link the answers to the repositories and replace by a code before the analysis. We received 33 responses, from all 18 teams (92\%). All quotes in this paper are our translations. The questionnaire and its English translation are available in our artifact folder (Section~\ref{sec:data}).

We wrote scripts that compute every measure below, available in our artifact folder. They rely on file names, trailers, and regular expressions, and depend on students having followed the conventions.

\subsection{RQ1: Adherence}

We checked each repository against the requirements and conventions: the URL in the \texttt{README.md}; dated specification files; whether each specification named in a \texttt{Spec:} trailer had been added in an earlier commit than the one naming it (\emph{Spec first}); recorded sessions; commits with both trailers; distinct human committers; lines of code with \texttt{cloc}~1.98 and the command of the assignment; and, although not required, automated tests. Some students committed with more than one e-mail address, such as a personal address and the no-reply address that GitHub assigns to commits made through its web interface; we merged the addresses of the same student, which affected 12 students in 10 projects. To estimate how complete the record is, we computed the share of commits attributed to an agent that fall within two hours of a recorded session.

\subsection{RQ2: Specifications}

We analyzed every Markdown file under \texttt{SPEC/} in the last commit of each repository along three dimensions. \emph{Authorship}: we classified a specification as written by an agent when the commit that added it carries an \texttt{Agent:} trailer naming a tool, a \texttt{Co-Authored-By} trailer naming an agent, or an agent as its author (\emph{By agent}); as written by a human when the trailer is \texttt{Agent: none}; and as undeclared otherwise. We also searched the sessions for prompts that ask the agent to write a specification. \emph{Structure}: we identified the sections on what and why, acceptance criteria, and what is out of scope by their headings, and counted as acceptance criteria the items listed under that section, as a list or a table, including subsections (\emph{AC}). \emph{Traceability}: we linked commits to specifications through the \texttt{Spec:} trailer, by file name or, when students named the part of the system instead (e.g., \texttt{Spec: triagem}, ``\emph{triage}''), by matching its words with the specifications present at that commit. For each specification, we computed whether a commit implements it (\emph{Referenced}), whether it was added in the same commit as code, that is, not written before it (\emph{With code}), and whether it was edited after its first implementing commit (\emph{Revised}); for each repository, the share of added lines that come from commits pointing to an existing specification (\emph{Traced}). Attribution based on trailers is common in studies of agent contributions~\cite{li2026aidev,zi2025agentpack}, but developers do not always declare AI-generated code~\cite{kashif2026selfdeclaration}, so undeclared commits may also have been produced by an agent.

\subsection{RQ3: Perceived Ownership}

We report the distribution of the answers and, in the text, the share of respondents that agree (4 or 5) or disagree (1 or 2) with each statement. We score each construct as the mean of its statements, after inverting the reverse-coded ones, and report its internal consistency with Cronbach's $\alpha$~\cite{cronbach1951}, which checks, for example, whether the seven statements on ownership of the code agree with each other enough to be averaged into one ownership score per student. Since Google Forms cannot require two numbers to add up to 100, we kept the shares of ideas and of code as reported (in 5 answers about ideas and 4 about code they add up to less than 100, probably leaving the remainder to the partner, and in 2 about code they exceed 100), and compare them with a Wilcoxon signed-rank test~\cite{wilcoxon1945}, to test whether students credit themselves with more of the ideas than of the code; it compares the two answers of each student without assuming that they follow a normal distribution. We coded the open answers following an open coding approach currently coded by one author and reviewed by a second author; disagreements were discussed until consensus as found.

\subsection{RQ4: Ownership and the Repository}

We relate the ownership constructs to the self-reported variables and to the data mined from the repository of each respondent, such as the share of commits produced by an agent, the share of specifications written by the agent, and the share of commits authored by the respondent, using Spearman's correlation~\cite{spearman1904}, to test, for example, whether students who read more of the generated code also report more ownership; it works on rankings, such as Likert answers, and with a sample as small as ours. Within pairs, we compare the member with more commits with the member with fewer with the same Wilcoxon test, to see whether the member who drove the agent more feels more ownership than their partner; since both worked on the same project, the comparison is not confounded by differences between projects. We also compare the share of the code each student attributes to themselves with the origin declared in their own commits, and read the open answers against the repository of each respondent. Given the sample size and the number of tests, these analyses are exploratory.

\subsection{RQ5: Requirement Gaming}
\label{sec:gaming-method}

While computing the line counts, we noticed that some files consisted of the same few lines repeated thousands of times. We then audited all 18 repositories for \emph{requirement gaming}, which we define as meeting the letter of a requirement without its purpose. Each check derives from one requirement and runs on every project, and every candidate was inspected manually by one author and reviewed by a second one. For the \emph{100,000 lines}, we flagged code that is counted but is not application: code files that no entry point reaches through imports; duplicated code, that is, lines in a sequence of 100 tokens that appears more than once in the project after names and numbers are replaced by placeholders; scripts that write source files from a loop; and test files that import no application code. We quantify the inflation of a project as the duplicated lines that the application never uses, plus the scripts that generated them. For \emph{specification first}, we looked for a forged record. Since rewritten commits replace the original ones, we queried GitHub's record of repository activity, which lists every force push and branch creation or deletion, and recovered the replaced history through GitHub's API; we also flagged specification commits whose timestamps have seconds equal to zero, against the rate of 1/60 expected by chance, file names dated differently from their commit, and specifications that declare they were written after the code. For the \emph{complete record}, we flagged commits that declare an agent on days without a recorded session, and searched the sessions for mentions of the size target (e.g., ``\emph{100k}'', ``\emph{cloc}''). For each confirmed case, we identified the commit that introduced it and searched the sessions for the prompts that led to it.

\emph{Ethical Considerations} 
This study followed the Brazilian guidelines for research with human participants (Resolution CNS 510/2016). Participation was voluntary: students agreed to take part, could withdraw at any time, and declining or withdrawing had no effect on their grades. 

\section{Participants and Projects}
\label{sec:participants}

\emph{The participants.} Of the 48 students enrolled in the course, 36 delivered the assignment, in 18 pairs, and 33 of them answered the survey (92\%), covering all 18 teams. Unless stated otherwise, percentages refer to the 33 respondents; questions that allowed several answers do not add up to 100\%.

\emph{Demographics and education.} The respondents are mostly young adults: 52\% are 18 to 25 years old and 39\% are 26 to 35, and three are older than 35. Gender was asked as an open question, and we coded the answers: twenty-four respondents (73\%) identify as men, seven (21\%) as women, and two did not report a gender. They are close to graduation: 82\% are in the seventh or eighth semester of the program, and 15\% in the ninth or later.

\emph{Programming and professional experience.} Half of the respondents (52\%) have been programming for three to five years, counting the time before college; 36\% for two years or less, and four for more than five years. Most (64\%) have worked in software development, mostly as interns (52\%).

\emph{Prior use of AI.} The respondents were not new to AI for programming. Before the assignment, 70\% used AI tools for programming every day or several times a day, and only four used them once a month or less. Almost all had used it to write code (94\%). Only 21\%, however, had let an agent perform tasks autonomously, which is what the assignment required, and 39\% had never written a specification before asking an agent for code; only 12\% had done so often.

\emph{Attitudes and self-efficacy.} The respondents value agents more than they find them easy: 76\% agree that agents are useful in their daily routine and 79\% that they increase their productivity, but only 39\% find them easy to use. Their confidence in directing agents is higher than their confidence in checking them: 61\% believe they can write instructions that lead the agent to the result they want and 58\% that they can notice when it deviates, but only 36\% believe they can assess whether the generated code is correct. Their self-efficacy in programming is moderate (mean of 3.18 on a five-point scale) and weakest precisely in the activities that verify code: 55\% do not believe they can debug an error without help, and 52\% that they can write tests. Only 39\% believe they can describe in writing what a system must do with verifiable criteria, the core skill of SDD.

\emph{The projects.} The 18 applications cover a wide range of domains: booking and scheduling for sports courts, beauty salons, medical clinics, two games, etc. Under the official count, the applications add up to 780,614 lines of code, with a median of 11,115 lines per project (min: 753, max: 201,848). The 18 repositories have 1,051 commits, 795 (76\%) of them attributed to an agent. Claude Code was the most used agent (10 projects), followed by Codex (5), Antigravity or Gemini (5), and GitHub Copilot (3); seven teams used more than one tool, and one team used five. Most are web applications written in TypeScript (9 projects, mostly with React or Next.js) or Python (7, mostly with FastAPI); one uses Java with Spring Boot and Angular, and one plain JavaScript.

% \begin{figure*}[t]
% \centering
% \includegraphics[width=\textwidth]{figures/participants.pdf}
% \caption{Demographics and experience of the 33 respondents. Gender was an open question, coded by us. Professional experience allowed several answers; the one respondent who selected both ``none'' and an internship was counted as having an internship.}
% \label{fig:participants}
% \end{figure*}

\section{Results}

\begin{table*}[t]
\centering
\caption{Projects, their adherence to the assignment (RQ1), and their specifications (RQ2). Agents: tools declared in commit trailers and in the README (AG Antigravity, CC Claude Code, Co GitHub Copilot, Cx Codex, DS DeepSeek, Ge Gemini, GPT ChatGPT, OC OpenCode). SLOC: lines of code counted with the command of the assignment; green marks the projects that reached 100,000 lines. URL: a public production URL is listed in the README. Specs: specification files whose name carries a date, over all specification files. Sessions: recorded session exports. Trailers: share of commits that carry both the Agent and the Spec trailers. Authors: distinct human committers, after merging the e-mails of the same student. Tests: automated test cases, not required by the assignment, excluding the generated tests of Trim. Spec first: share of the specifications named in Spec trailers that had been added in an earlier commit than the one naming them. By agent: share of specifications added by a commit attributed to an agent. AC: share of specifications with at least one acceptance criterion. Referenced: share of specifications that at least one commit points to. With code: share of specifications added in the same commit as code, that is, not written before it. Revised: share of specifications edited after the first commit that implements them. Traced: share of the added lines of code that come from commits pointing to an existing specification. Bold: fails a requirement.}
\label{tab:projects}
\scriptsize
\renewcommand{\arraystretch}{1.08}
\setlength{\tabcolsep}{2pt}
\begin{tabular*}{\textwidth}{@{\extracolsep{\fill}}r l l r c r r r r r r r r r r r r@{}}
\toprule
& & & & \multicolumn{6}{c}{\textbf{Process (RQ1)}} & \multicolumn{7}{c}{\textbf{Specifications (RQ2)}} \\
\cmidrule(lr){5-10} \cmidrule(l){11-17}
\textbf{ID} & \textbf{Project} & \textbf{Agents} & \textbf{SLOC} & \textbf{URL} & \textbf{Specs} & \textbf{Sessions} & \textbf{Trailers} & \textbf{Authors} & \textbf{Tests} & \textbf{Spec first} & \textbf{By agent} & \textbf{AC} & \textbf{Referenced} & \textbf{With code} & \textbf{Revised} & \textbf{Traced} \\
\midrule
1 & \textsf{BookIt} & AG, CC & 26,386 & \checkmark & 15/15 & 14 & 49\% & 2 & 339 & 56\% & 87\% & 47\% & 93\% & 0\% & 13\% & 97\% \\
2 & \textsf{EduQuest} & OC, CC & 12,580 & \checkmark & 13/13 & 1 & 32\% & 2 & 201 & 58\% & 85\% & 85\% & 92\% & 0\% & 46\% & 62\% \\
3 & \textsf{Trim} & AG & \cellcolor{green!25}201,848 & \checkmark & 14/14 & \textbf{0} & 100\% & \textbf{1} & 0 & 53\% & 100\% & 100\% & 100\% & 7\% & 0\% & 2\% \\
4 & \textsf{KapiTrace} & Cx, AG & 4,884 & \checkmark & 17/17 & 24 & 64\% & 2 & 85 & 78\% & 24\% & 6\%$^{\dagger}$ & 24\% & 76\% & 12\% & 24\% \\
5 & \textsf{GameLog} & CC, Co & 4,084 & \checkmark & 18/18 & 4 & 82\% & 2 & 161 & 87\% & 44\% & 100\% & 100\% & 0\% & 17\% & 96\% \\
6 & \textsf{SalonBook} & Cx, DS & 20,516 & \checkmark & 25/25 & 27 & 97\% & 2 & 278 & 72\% & 92\% & 96\% & 100\% & 0\% & 12\% & 87\% \\
7 & \textsf{ServiceFlow} & Co & 9,281 & \checkmark & 2/2 & 17 & 83\% & \textbf{1} & 126 & 88\% & 0\% & 100\% & 100\% & 0\% & 50\% & 87\% \\
8 & \textsf{MediaPlay}$^{\S}$ & AG & 1,262 & \checkmark & \textbf{0/4} & \textbf{0} & \textbf{0\%} & 2 & 0 & -- & 0\% & 100\% & 0\% & 100\% & 0\% & 0\% \\
9 & \textsf{Finances} & CC & 35,597 & \checkmark & 52/52 & 51 & 66\% & 2 & 883 & 62\% & 100\% & 100\% & 98\% & 13\% & 8\% & 92\% \\
10 & \textsf{HealthOS} & CC & 20,850 & \checkmark & 23/23 & 1 & \textbf{0\%} & 2 & 193 & --$^{*}$ & 65\% & 100\% & 43\% & 0\% & 4\% & 0\% \\
11 & \textsf{DeliveryFast} & GPT, Ge & \cellcolor{green!25}137,733 & \checkmark & 3/3 & 1 & 92\% & 2 & 0 & 84\% & 67\% & 100\% & 100\% & 0\% & 0\% & 100\% \\
12 & \textsf{ClearTable} & Cx & 753 & \checkmark & 2/2 & 1 & 90\% & \textbf{0}$^{\ddagger}$ & 0 & 78\% & 100\% & 100\% & 100\% & 0\% & 0\% & 100\% \\
13 & \textsf{CourtAuctions} & CC & 9,650 & \checkmark & 4/6 & \textbf{0} & \textbf{0\%} & \textbf{1} & 259 & -- & 0\% & 0\% & 0\% & 100\% & 0\% & 0\% \\
14 & \textsf{ElectionOps} & Cx, CC, DS, Co, GPT & 79,741 & \textbf{$\times$} & 28/29 & 17 & 65\% & 2 & 673 & 83\% & 52\% & 83\% & 62\% & 31\% & 3\% & 71\% \\
15 & \textsf{CircuitCrash} & CC & 3,902 & \checkmark & 5/5 & 7 & 72\% & 2 & 138 & 81\% & 100\% & 100\% & 60\% & 0\% & 40\% & 30\% \\
16 & \textsf{RightMovie} & CC & 9,442 & \checkmark & 14/14 & 4 & 97\% & \textbf{1} & 281 & 60\% & 100\% & 100\% & 100\% & 0\% & 50\% & 100\% \\
17 & \textsf{UmbraTrail}$^{\P}$ & Cx & \cellcolor{green!25}196,302 & \checkmark & 22/23 & 20 & \textbf{5\%} & 2 & 150 & 75\% & 4\% & 96\% & 4\% & 0\% & 4\% & 0\% \\
18 & \textsf{ClinicBooking} & CC & 5,803 & \checkmark & 12/12 & 2 & 91\% & 2 & 129 & 90\% & 100\% & 100\% & 58\% & 0\% & 0\% & 99\% \\
\midrule
 & \textbf{Total} &  & \textbf{780,614} &  & \textbf{269/277} & \textbf{191} &  &  & \textbf{3,896} &  & \textbf{69\%} & \textbf{86\%} & \textbf{71\%} & \textbf{14\%} & \textbf{12\%} & \textbf{39\%} \\
\bottomrule
\end{tabular*}

\vspace{2pt}
{\scriptsize\raggedright $^{\S}$OpenSpec layout, undated; acceptance criteria are \emph{WHEN}/\emph{THEN} scenarios. $^{\P}$Rewritten history; the original one has no specifications (Section~\ref{sec:gaming}). $^{*}$\texttt{Spec:} trailers name the part of the system, not a file. $^{\ddagger}$All commits authored by Codex. $^{\dagger}$13 of the 17 specifications are eight-word placeholders.\par}

\end{table*}

\subsection{RQ1: Do students follow the rules of the assignment?}

Table~\ref{tab:projects} summarizes the adherence of the 18 projects to the methodology. Seventeen projects listed a public URL, and 17 had a \texttt{SPEC/} folder, almost always following the dated naming convention; \textsf{MediaPlay} kept its four specifications elsewhere: the team chose OpenSpec~\cite{openspec}, an open-source SDD framework not required by the assignment, which keeps one specification per capability under \nolinkurl{openspec/changes/<change>/specs/<capability>/spec.md} and expresses acceptance criteria as \emph{WHEN}/\emph{THEN} scenarios. We include these four specifications in the analysis of RQ2, counting each scenario as one acceptance criterion. Writing the specification before the code was followed only in part: among the 15 projects in which we could compute it, between 53\% and 90\% of the specifications named in \texttt{Spec:} trailers had been added in an earlier commit (column \emph{Spec first} of Table~\ref{tab:projects}). Recording sessions was the most neglected rule: three projects recorded no session at all, and four recorded a single one. Even where sessions were recorded, they cover only part of the work: 48\% of the 795 commits attributed to an agent fall within two hours of a recorded session. In the sessions we could parse, students wrote 741 prompts, with a median of 15 words; 13\% have at most three words, such as ``\emph{continue}''. The median share of commits with both \texttt{Agent:} and \texttt{Spec:} trailers is 69\%, but three projects have no commit with both. Four projects have a single human committer (\textsf{Trim}, \textsf{ServiceFlow}, \textsf{CourtAuctions}, and \textsf{RightMovie}), and in \textsf{ClearTable} all 10 commits were authored and committed by Codex alone, under the identity \texttt{Codex <codex@local>} and with no co-author trailer, so the repository does not show which student drove each session. Although the assignment did not require tests, 14 projects include them, with 3,896 test cases that make up 15\% of their code once the generated tests of \textsf{Trim} are excluded (Section~\ref{sec:gaming}), and four run them in continuous integration.

\subsection{RQ2: How do students specify software for AI coding agents?}

Table~\ref{tab:projects} also summarizes, per project, the 277 specifications found in the 18 repositories. Together, they add up to 261,353 words, with a median of 544 words per specification.

\emph{RQ2.1: Who writes the specification?} The agent, in most cases. Of the 277 specifications, 191 (69\%) were added by commits attributed to an agent (column \emph{By agent} of Table~\ref{tab:projects}), only 3 (1\%) by commits explicitly declared as handwritten (\texttt{Agent: none}), and 83 (30\%) by commits that do not declare their origin. Undeclared does not mean handwritten: the four specifications of \textsf{MediaPlay}, for instance, were committed without any trailer, but the prompts recorded by the team show that they were generated by the agent from a one-paragraph description. In six projects (\textsf{Trim}, \textsf{Finances}, \textsf{ClearTable}, \textsf{CircuitCrash}, \textsf{RightMovie}, and \textsf{ClinicBooking}), every specification was committed by an agent. Specifications committed by an agent are also the most complete: 91\% of them have acceptance criteria, against 73\% of the undeclared ones, and they are longer (median of 677 against 197 words) and less often left without any implementing commit (12\% against 69\%). The artifact SDD assigns to the human was, in most projects, also written by the agent. The sessions show how little some of these specifications required: we found 27 prompts, in 8 projects, that ask the agent to write a specification, with a median of 120 words, from four words (``\emph{you can write the spec}'') to detailed instructions that dictate its sections; in \textsf{Finances}, a single prompt of 66 words asked for ``\emph{a spec in SPEC/ for each part}'' of the application. This is not necessarily a deviation from how SDD is promoted: in GitHub Spec Kit, the coding agent drafts the specification from a high-level description, and the developer reviews it. Whether reviewing a specification gives students the same sense of ownership as writing it is one of the questions our survey addresses.

\emph{RQ2.2: Do specifications follow the required structure?} Mostly. 91\% of the specifications state what is built and why, 85\% state what is out of scope, and 86\% list acceptance criteria (column \emph{AC} of Table~\ref{tab:projects}), with 3,102 criteria in total. The exceptions are concentrated: \textsf{BookIt} lists acceptance criteria in fewer than half of its specifications; \textsf{CourtAuctions} writes long technical documents organized by layer (stack, layout, pages, tests) rather than by feature; and 13 of the 17 specifications of \textsf{KapiTrace} are placeholders, with the title and the sentence ``\emph{Details will be added here}'', committed on the first day and never filled in.

\emph{RQ2.3: Do specifications govern the code?} Partially. 71\% of the specifications are referenced by at least one commit (column \emph{Referenced} of Table~\ref{tab:projects}), and only two were referenced by commits made before they existed. However, 14\% of the specifications were added in the same commit as the code they describe (column \emph{With code} of Table~\ref{tab:projects}), which is the case for all specifications of \textsf{CourtAuctions} and \textsf{MediaPlay} and most of those of \textsf{KapiTrace}. Only 12\% of the specifications were edited after their implementation started (column \emph{Revised} of Table~\ref{tab:projects}), and only one explicitly replaces an earlier one, which suggests that specifications were written once, before or along with a part of the system, and rarely revisited as the system evolved. Finally, only 39\% of the lines added to the repositories come from commits that point to an existing specification (column \emph{Traced} of Table~\ref{tab:projects}). Projects that followed the trailer convention consistently, such as \textsf{ClinicBooking}, \textsf{DeliveryFast}, \textsf{RightMovie}, \textsf{GameLog}, and \textsf{Finances}, trace more than 90\% of their lines to a specification, whereas \textsf{Trim}, with 63 commits marked \texttt{Spec: none}, traces only 2\%, and \textsf{UmbraTrail}, whose commits rarely carry a \texttt{Spec:} trailer, traces none.

\subsection{RQ3: How much ownership do students perceive?}

Section~\ref{sec:participants} describes the respondents, and Figure~\ref{fig:survey} shows the distribution of their answers to all statements.

\emph{Ownership of the code is divided.} Ownership of the code has a mean of 3.29 (SD 1.14), and the answers spread over the whole scale. Only 27\% of the students agree that the code is truly theirs, and 48\% disagree. Yet 52\% agree that it reflects their own ideas, 52\% that they chose the main decisions it embodies, and 52\% would put the project in their portfolio. Five students agree that the code reflects their ideas while disagreeing that it is theirs, and four would put in their portfolio a project they do not feel is theirs, a pattern reminiscent of the AI ghostwriter effect~\cite{draxler2024}.

\emph{The ideas are mine; the code is the agent's.} The percentages make this split explicit. Students attribute to themselves a median of 50\% of the ideas of the project (Q1: 30\%, Q3: 80\%), but only 20\% of the final code (Q1: 5\%, Q3: 30\%), a significant difference (Wilcoxon signed-rank test, $p < 0.001$). 14 students (42\%) attribute at most 10\% of the code to themselves, and 5 attribute none. Students also read little of what the agent wrote: the median share of the generated code read before committing is 30\%, and 15 students (45\%) read at most a fifth of it. The median time dedicated to the assignment was 8 hours (Q1: 5, Q3: 14).

\emph{Owning the specification.} Ownership of the specification and the application has a mean of 3.27. Only 33\% of the students agree that the specification is truly theirs, which is consistent with the agent having committed most specifications (RQ2). About half agree that the specification describes the system they wanted (48\%) and that they noticed when the agent deviated from it (48\%).

\emph{Responsibility without knowledge.} Knowledge of the code is the lowest construct, with a mean of 2.71. Most students could not explain any part of the code to a colleague (58\% disagree), do not know where each feature is implemented (61\%), and could not modify the code without the agent (55\%). Still, 52\% feel responsible for its defects; of the 17 students who do, six could not modify the code without the agent.

\emph{What students say.} Thirty students answered at least one open question. The most frequent theme is a division of labor in which the ideas, the design, and the decisions belong to the student and the code belongs to the agent: ``\emph{I was the engineer of the construction, and the AI was the bricklayer; I designed and supervised. I cannot say it is totally and truly mine.}'' Some students extend authorship to whoever had the idea (``\emph{even if it executes the code, authorship belongs to whoever had the idea and formulated it}''), while others tie it to writing the code (``\emph{I would only truly feel that the code was mine if I wrote it 100\%}''). Several answers attribute the loss of control or of review to the pressure of the size target and the deadline: ``\emph{due to the stress of the 100 thousand lines and little time, I did not dedicate much time to reviewing the code generated by the agent, [\ldots] the project was slipping out of my control.}'' Others describe accepting code without reading it, validating only the interface, the tests, or a playtest: ``\emph{if it was working, I simply accepted it, without analyzing the code itself. [\ldots] As much as I feel I built the project, I know I do not really master the code.}'' Some report the agent drifting from what they asked, sometimes unnoticed: ``\emph{the tests kept passing and the features seemed to be progressing, but [\ldots] there were still requirements of the specification that had not been implemented.}'' A few describe strategies to stay in control, such as approving every commit, using a plugin that makes the agent plan and ask before deciding, writing part of the code by hand, or following SDD, which ``\emph{helped me have more authority over the project.}'' Finally, some students raised the cost of agent subscriptions and token limits, which one of them considered unfair for students of a public university.

\begin{table}[t]
\centering
\caption{Spearman correlations between perceived ownership and self-reported and mined variables ($n = 33$). $^{*}p < 0.05$; $^{**}p < 0.01$; no correction for multiple comparisons.}
\label{tab:corr}
\footnotesize
\setlength{\tabcolsep}{3pt}
\begin{tabular}{@{}l r r r@{}}
\toprule
& \textbf{Own.} & \textbf{Own.} & \textbf{Know-} \\
& \textbf{code} & \textbf{spec/app} & \textbf{ledge} \\
\midrule
\% of ideas attributed to self & $+$0.37$^{*}$ & $+$0.36$^{*}$ & $+$0.22 \\
\% of code attributed to self & $+$0.44$^{*}$ & $+$0.21 & $+$0.49$^{**}$ \\
\% of agent code read & $+$0.43$^{*}$ & $+$0.36$^{*}$ & $+$0.35$^{*}$ \\
Hours dedicated & $+$0.28 & $+$0.23 & $+$0.25 \\
Self-efficacy in programming & $+$0.57$^{**}$ & $+$0.38$^{*}$ & $+$0.60$^{**}$ \\
Self-efficacy with AI tools & $+$0.45$^{**}$ & $+$0.39$^{*}$ & $+$0.48$^{**}$ \\
Acceptance of AI coding agents & $+$0.50$^{**}$ & $+$0.44$^{*}$ & $+$0.35$^{*}$ \\
\midrule
Own share of the team's commits & $+$0.12 & $+$0.17 & $+$0.11 \\
Team: commits by agent & $-$0.18 & $-$0.03 & $-$0.04 \\
Team: specifications by agent & $-$0.14 & $-$0.13 & $-$0.14 \\
Team: recorded sessions & $+$0.10 & $-$0.12 & $+$0.10 \\
Team: lines traced to a specification & $-$0.01 & $+$0.07 & $+$0.07 \\
\bottomrule
\end{tabular}
\end{table}

\subsection{RQ4: How does ownership relate to the agent's recorded involvement?}

Table~\ref{tab:corr} shows the Spearman correlations between the three ownership constructs and the other variables. Ownership is associated with what students report about their own involvement: the share of the code they attribute to themselves, the share of the generated code they read, and, to a lesser extent, the share of the ideas. It is also associated with their self-efficacy in programming and with AI tools, and with their acceptance of agents. In contrast, none of the measures mined from the repositories correlates with any ownership construct: neither the share of commits attributed to an agent, nor the share of specifications written by the agent, nor the number of recorded sessions, nor the share of lines traced to a specification, nor the share of commits authored by the respondent.

One reason may be that these measures describe the team, not the student: both members of a pair receive the same value, and almost every team delegated most of the code. Within teams, the picture changes. In the 14 pairs in which both members answered and their shares of commits differ, the member with more commits reported higher ownership of the code in 10 pairs (median of 3.64 against 2.71; Wilcoxon signed-rank test, $p = 0.046$), but not higher knowledge of the code ($p = 0.37$). Driving the agent, it seems, makes students feel the code is more theirs, without making them know it better.

\emph{Do self-reports match the record?} Commit trailers also declare whether a student had a hand in the code (\texttt{Agent: none} or \texttt{+ manual}). For the 15 respondents whose own commits declare their origin, the share of the code they attribute to themselves is unrelated to the share of those commits with a declared hand in the code ($\rho = 0.02$). Eleven of them never declared a hand in their own commits, yet attribute to themselves a median of 20\% of the code, and six of them at least 20\%. The remaining respondents have no commits of their own (6) or only commits that do not declare their origin (12).

\emph{Ideas without the artifact that records them.} Reading the open answers against the repositories sharpens the split of RQ3. A student of \textsf{RightMovie} feels the code is theirs ``\emph{because I am the one who has the ideas, chooses the technologies, and tries to guide the AI}'', attributes to themselves 70\% of the ideas and none of the code, and read none of it; every specification of the project, where SDD expects those ideas to be written, was committed by the agent. A student of \textsf{HealthOS} attributes to themselves 55\% of the code because ``\emph{the ideas, structure, and modeling were created before going to the agent}'', but none of their 46 commits declares a hand in the code. And the student of \textsf{UmbraTrail} who described themselves as ``\emph{the engineer of the construction}'' placed the moment they lost control ``\emph{when I really needed to increase the volume of code}'', in the team that generated 57 thousand lines of filler (Section~\ref{sec:gaming}). Students locate their ownership in ideas and decisions, but the artifact in which SDD records ideas and decisions was, for most of them, written by the agent.

\subsection{RQ5: Do students game the requirements, and what role do AI agents play?}
\label{sec:gaming}

Only three projects reached the required 100,000 lines: \textsf{Trim} (201,848), \textsf{UmbraTrail} (196,302), and \textsf{DeliveryFast} (137,733), which together account for 69\% of all the code delivered. They are exactly the three in which we found generated code that the application never uses, and one of them also rewrote its history. The audit flagged candidates in other projects, such as unreachable code files (in 17 projects), tests that import no application code (11), and generator scripts (6), but none of them survived manual inspection: they were, for instance, framework entry points, database migrations, and tests that exercise the application through its HTTP interface. Specification commits with seconds equal to zero appear in four projects, but only in \textsf{UmbraTrail} (13 of 22) are they unlikely by chance.

\emph{\textsf{DeliveryFast}: stub functions.} A script, \texttt{gerador.py} (``\emph{generator}''), writes a 24,999-entry constant array and seven files with 3,999 functions each, such as \texttt{renderModule\_cart\_400()}, that return a fixed HTML string. With the script, they account for 136,991 of the 137,733 lines of the project (99.5\%), and the application references none of them; its actual code has 742 lines. The second of the three commits that added them is titled ``\emph{generates volumetric code to reach the cloc target}''. The recorded session shows how it began: the student asked ``\emph{I need 100k lines of code. What do I have to do?}'', and the agent answered that ``\emph{since writing this volume by hand would be infeasible, one possibility is to use automated generation}'', wrote the script, and, when asked again, scaled it from 1,500 to 4,000 functions per file.

\emph{\textsf{Trim}: tests that test nothing.} A script writes 13 test files with 1,000 tests each, such as \texttt{test\_walkin\_scenario\_179}, which assert properties of a copy of a fixture (e.g., \texttt{assert 179 <= 1000}) and import no application code. With the script, they account for 97\% of the lines of the project (195,137 of 201,848); without them, it has 6,711. The single commit that added them, declaring Antigravity, reads ``\emph{absurd increase in the metrics, reaching 200 thousand lines in the project}''. The team recorded no session.

\emph{\textsf{UmbraTrail}: generated content and a rewritten history.} A script writes 920 content files cloned from a few templates (e.g., \texttt{ShrineProfilePack1.ts} to \texttt{ShrineProfilePack150.ts}), which no file of the game imports: 57,040 lines (29\% of the count), plus 10,810 lines of generator scripts that embed source code also written to \texttt{src/}. Their commit was originally titled ``\emph{generated content packs (\textasciitilde104k loc)}'', a day after the agent had warned the student that numbered files created ``\emph{only to reach 100 thousand lines would be artificial inflation to deceive the academic metric}''. The team also forged its process record. On October 6, the student asked the agent for specifications ``\emph{created from the first day of commit in the repository; it cannot be from today}''. That afternoon, GitHub's activity record shows a backup branch (\nolinkurl{backup/main-antes-specs-20261006}, ``\emph{main before specs}''), two force pushes, and the deletion of the branch. The original history, which GitHub still serves, has 39 commits and no specification; the delivered one has 75, of which 22 add specifications, three of them dated before the repository existed and 13 with timestamps whose seconds are zero ($p < 10^{-17}$ if seconds were uniform). Its measures in Table~\ref{tab:projects} therefore describe a fabricated history.

\emph{The role of AI agents.} The incentive to game the target was the students', and so were the decisions to commit the filler and to rewrite the history. The role of the agents varied. In \textsf{DeliveryFast}, the only case in which the session that produced the filler was recorded, the agent proposed the generator and scaled it up, without noting that the code would never be used; in \textsf{UmbraTrail}, the agent warned against inflating the count, to no effect. Prompts in two other projects asked the agent to write code ``\emph{in more lines}'' (\textsf{BookIt}) or to add ``\emph{rich business methods}'' ``\emph{to reach the 100k lines}'' (\textsf{HealthOS}), without producing templated or unused code, while \textsf{ElectionOps} told its agent that the target ``\emph{must be a consequence of the functional growth of the platform, and not of artificial code}''. Agents did not create the incentive to game the target; they lowered its cost to a prompt.

\section{Discussion}

\subsection{Agents Make Gaming Cheap}

Inflating a line count by hand takes effort, and its result is easy to recognize. With an agent, it takes one prompt, and the result compiles, deploys, and passes its own tests. RQ5 shows that, when a requirement can be satisfied without the engineering it is meant to elicit, some students will satisfy it that way, and agents will help them, and may even suggest how. The situation resembles \emph{specification gaming} in machine learning, in which a system satisfies the literal objective rather than the intended one. Here, however, the optimization was done by students, who had the incentive of a grade, and the agent served as a tool that made the literal objective easy to reach.

Three observations follow. First, every case was visible in the record: commit messages announce the filler, sessions show how it was produced, trailers name the agent, and GitHub's activity log exposes the rewritten history. The complete record that the assignment required, rather than the line count, is what made the gaming detectable, and it can be checked automatically. Second, the record itself can be forged: the same agent that writes code can write backdated specifications and rewrite history, so rules that live only in the repository need a source of truth outside it, such as the activity log of the hosting platform or continuous integration. Third, without the filler, only one project still reaches 100,000 lines, and the total delivered falls by about half, from 781 to roughly 380 thousand lines. The target did not measure what we wanted it to measure, and its failure is itself a finding about teaching with agents.

\subsection{Implications for Educators}

\emph{Size targets measure what the agent is good at.} The 100,000-line target was meant to make agents indispensable, and it did: no team could approach it by hand. But lines are the cheapest thing an agent produces, and the target was met unevenly: three projects produced 69\% of the code, while the median project stayed at 11,115 lines. A line count says little about the engineering behind it, and, as Section~\ref{sec:gaming} shows, it can be met with code that does nothing. Educators who want students to work at scale may set targets in terms of scope instead, such as the number of features whose acceptance criteria are demonstrably met, which also rewards the specification work the course is meant to teach.

\emph{Decide, explicitly, who writes the specification.} Our assignment required specifications to precede the code, but not that students write them, and 69\% were committed by an agent. If specifying is a learning goal, it has to be observable. A course may require a first draft of each specification committed by a student (\texttt{Agent: none}) before any agent edits it, and assess the difference between the student's draft and the final version; or, accepting that the agent drafts, assess the review instead, by asking students to record what they changed in the agent's draft and why. In both cases, an oral defense of a few acceptance criteria is a cheap check that students understand what they asked for.

\emph{Move process rules from honor to tooling.} The rules that depended on students' discipline eroded first: three teams recorded no session, one in seven specifications entered the repository with the code, and only 39\% of the added lines can be traced to a specification. Since agents write most commits, they can be instructed, for instance through \texttt{AGENTS.md}, to add the \texttt{Agent:} and \texttt{Spec:} trailers and to export their sessions, and a continuous integration check can reject commits without a specification or specifications added together with code. Automating the record also makes grading at this scale feasible.

\emph{Make specifications change.} Only 12\% of the specifications were revised after their implementation started, and only one replaced an earlier one. Specifications written once and never revisited teach little about how specifications work in practice. Introducing a change of requirements midway through the assignment would force students to write a new specification, decide what it replaces, and see whether the agent follows it.

\emph{Assess ownership, not only output.} A published application that opens with one click shows that the agent worked; it does not show that the students can explain, change, or defend it. In our survey, 58\% of the students could not explain any part of their code to a colleague, 55\% could not modify it without the agent, and 45\% read at most a fifth of the generated code before committing it; and the one variable under students' control that we found associated with ownership was how much of that code they read. Short oral checks, in which students locate a feature in the code or explain a decision recorded in a specification, are a practical complement to the presentation.

\subsection{Threats to Validity}

\emph{Construct validity.} Attribution to an agent relies on commit trailers, co-author lines, and bot authors; commits without them may also have been produced by an agent~\cite{kashif2026selfdeclaration}, so our counts are lower bounds. Whether a specification preceded the code that names it depends on \texttt{Spec:} trailers that point to a file, so we could not compute it for three projects: \textsf{MediaPlay} and \textsf{CourtAuctions} have no commit with a \texttt{Spec:} trailer, and the trailers of \textsf{HealthOS} name the part of the system in free text (e.g., \texttt{spec: triagem}) rather than a file. For these projects, only whether each specification was added together with code (RQ2) tells the order of specification and code. The structure of specifications was assessed through section headings and list items, which may miss sections whose headings we did not anticipate. Ownership is self-reported, and the origin declared in commit trailers is coarse: \texttt{+ manual} does not say how much was edited by hand. Session exports vary by tool, 22 of 109 could not be parsed, and some are summaries rather than transcripts, so the prompt counts are lower bounds. \emph{Internal validity.} Students knew that the record was part of the grade, which may have shaped what they declared; the survey was not graded, but students answered it for their instructor, which may induce socially desirable answers. Our search for requirement gaming relies on heuristics that detect templated and cloned code; filler that varies more, such as verbose but distinct functions, would escape it, so the cases we report are a lower bound. Detecting rewritten histories depends on GitHub's activity record. \emph{External validity.} We study one elective course, at one university, with 18 teams, under an ambitious size target; students who chose the course may be more enthusiastic about AI than their peers, and the tools evolve quickly. \emph{Conclusion validity.} With 18 teams, the relations we report in RQ4 are exploratory.

\section{Related Work}

\emph{AI in software engineering courses.} Most evidence on how software engineering students use generative AI comes from surveys and experience reports. Students use it mostly to implement and to learn, value it for brainstorming, and struggle to adapt its output~\cite{qin2026role,baresi2025students}; in capstones and team projects, it reshapes roles and calls for team-level guidelines~\cite{mircea2026capstone,kharrufa2026llms,salomon2025exploration}. Courses now integrate it deliberately, assessing process instead of output~\cite{fein2026aidriven}, moving from specification to implementation~\cite{wang2026specifications}, or adopting SDD~\cite{martin2026aisdlc,geng2026mapping}. These studies cover longer projects, external clients, and instructional design, which we do not. But they mostly rely on what students report, in settings where AI supports parts of the work; Rasnayaka et al. are an exception, tracing generated code and prompts in a course project~\cite{rasnayaka2024empirical}. We study a setting in which the agent does almost all of the work, and check what students report against what their repositories record. The courses that adopt SDD assume that students asked to specify first will be the ones specifying; none of them examines who wrote the specifications.

\emph{How students work with AI.} Studies of student--AI interaction show that many students delegate rather than refine: most students in an introductory course submitted AI-generated code almost unchanged~\cite{shao2026trajectories}, novices ask an assistant for a summary instead of asking it to clarify the code~\cite{mohamed2026beyond}, and students who vibe code rarely read the output~\cite{geng2026vibe,gama2026vibes}. These studies observe short tasks and, unlike ours, measure comprehension or learning directly. We observe delegation one level up, in the specification that SDD intends as the human's contribution, and at a scale where reading all the generated code is no longer possible.

\emph{Ownership.} Psychological ownership, the feeling that something is ``\emph{mine}''~\cite{pierce2001}, differs from the structural ownership that software engineering measures by contribution~\cite{bird2011dont}. In writing, people do not feel they own AI-generated text yet claim its authorship~\cite{draxler2024}, and ownership drops most when the AI drafts the text~\cite{gero2026planning,joshi2025writing}; in programming, it drops with the share and autonomy of the AI~\cite{bergh2026cognition,seo2026whose} and depends on how much students refine the output~\cite{ye2026attribution}, while professionals keep their agency by delegating in detail~\cite{feng2026junior,huang2025dontvibe}. These studies isolate causes in controlled tasks, which an observational study cannot. We test whether the same dissociation holds for systems students built and presented, and find that self-reports are disconnected not only from feelings of ownership but also from the record of who wrote what.

\emph{Specifications, integrity, and agent traces.} Spec-driven tools make the specification the source of truth~\cite{piskala2026sdd}, in line with research on making intent explicit~\cite{fakhoury2024ticoder,lahiri2026intent}, but also let the agent draft it; we show what happens to this artifact when students use these tools. Students report misusing AI mostly in programming assignments under time pressure~\cite{desouzasantos2026cheating}; we document, in the artifacts themselves, how a metric was gamed and what the agent contributed. Finally, we identify agent work through trailers and bot authors, as studies of professional repositories do~\cite{li2026aidev,zi2025agentpack}, and combine it with self-reports, since perceived and measured effects of AI can diverge~\cite{becker2025measuring}.

\section{Conclusion}

We asked students to build, with AI coding agents and under SDD, systems larger than they could write by hand, and recorded how they did it. The agents wrote not only the code but most of the specifications meant to keep students in control; students located their ownership in ideas they rarely wrote down, could seldom explain or modify the code, and reported contributions their commits do not show; and the size target was met with code that does nothing. Agents did not create these problems, but they made each of them cheap. Courses that adopt agents should therefore grade what agents do not provide for free, namely the specification students write, the review they do, and the code they can explain, and enforce the engineering process with tools rather than trust.

\subsection{Future Work}

We plan to evaluate these changes in the next offering of the course, requiring students to draft their own specifications, setting targets by scope rather than by size, and enforcing the conventions with continuous integration. We also plan to follow the same students through the semester project and its individual oral defenses, to test whether perceived ownership predicts whether students can explain and change their code, and to analyze in depth how they write prompts, the one record of their decisions that the agent does not write.

\section*{Data Availability}
\label{sec:data}

The replication package, available at \url{https://anonymous.4open.science/r/code-ownership-ai-3ACA/}, contains the scripts that compute every number, table, and figure in this paper, the version of \texttt{cloc} we used (1.98), the inputs we coded manually, and the anonymized survey responses, from which the scripts reproduce every survey result except the age and gender of the participants. The team repositories, session transcripts, open survey answers, and the mapping between students and their commits contain personal data and are not shared.

\bibliographystyle{IEEEtran}
\bibliography{refs}

\end{document}